\documentclass[twocolumn,showpacs]{revtex4}    

\usepackage{graphicx}

\begin{document}

\title{Switching of optical vortices in nonlinear annular couplers}

\author{Jos\'e R. Salgueiro}

\address{Departamento de F\'{\i}sica Aplicada, Universidade de Vigo,\\
Facultade de Ciencias de Ourense, E-32004 Ourense, Spain}

\author{Yuri S. Kivshar}

\address{Nonlinear Physics Center, Research School of Physical
Sciences and Engineering, Australian National University, ACT 0200 Canberra,
Australia}

\begin{abstract}
We suggest an annular waveguide coupler for switching the angular
momentum of light. We study linear and nonlinear coupling of both
power and momentum of an optical vortex beam launched into one of
the ring cores, and demonstrate that the switching takes place well
below the collapse threshold. The switching is more effective for
the inner-ring excitation since it triggers more sharply and for the
lower power enough to avoid the beam azimuthal instability.
\end{abstract}


\maketitle

Optical vortices are fundamental structures in the light fields
associated with the points of vanishing intensity and phase
singularities of optical beams~\cite{review}. Optical vortices are
generated experimentally in different types of linear and nonlinear
optical media~\cite{review,PIO}. However, when a vortex beam
propagates in a nonlinear medium, it becomes unstable due to the
symmetry-breaking azimuthal instability~\cite{PIO}, and it decays
into several fundamental solitons~\cite{book}. Therefore, it is
commonly accepted that any kind of an optical device operating with
a transfer of the angular momentum of light and nonlinear switching
of an optical vortex would be impossible due to this inherent
instability and the subsequent vortex decay.

In a contrast with this common belief, in this Letter we suggest a
novel type of a nonlinear waveguide coupler composed by two weakly
coupled ring waveguides which not only preserve the angular momentum
of the input light during the propagation, but also allow nonlinear
switching for the beam power and angular momentum when it operates
with ring-like optical vortex beams. Such two-ring annular waveguide
couplers can be created, in particular, by a proper modulation of
nondiffracting ring Bessel-like optical
lattices~\cite{Kartashov:2004-93904:PRL} recently generated
experimentally by the optical induction technique~\cite{zhigang}. In
this context, the ring solitons were studied in a multi-ring lattice
for the case of a saturable nonlinearity~\cite{Hoq2005}.

The purpose of this Letter is twofold. First, we suggest a simple
design of the nonlinear vortex coupler that is consistent with the
conservation of the angular momentum of the vortex beam.  Second, we
study the angular momentum switching of the vortex in this coupler
and compare it with the familiar switching observed for the beam
power. We are interested how an input beam with a nonzero angular
momentum can tunnel between the ring-like cores (either from the
outer ring or from the inner ring) of the annular coupler, in both
linear and nonlinear regimes. To the best of our knowledge, this
problem has never been addressed before, but it seems very important
for suggesting novel ways to manipulate, transform, and control the
angular momentum of light.
\begin{figure}[htbp]
  \centerline{
    \scalebox{0.28}{
      \includegraphics{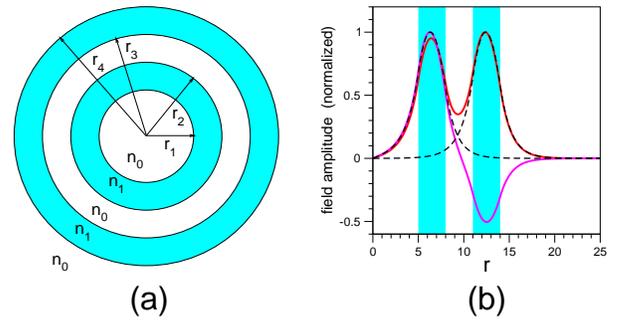}}
  }
\caption{(a) Sketch of an annular coupler. Values used for all
simulations are $r_1=5$, $r_2=8$, $r_3=11$, $r_4=14$, $n_0=0$,
$n_1=1$. (b) Continuous lines: radial profiles of the linear
lowest-order vortex-type modes of the coupler with winding number
$\ell=1$ (symmetric, $\beta=0.637$, and antisymmetric,
$\beta=0.594$). Dashed lines: radial profiles of the lowest-order
vortex-type mode of each separate waveguide (inner: $\beta=0.608$,
outer: $\beta=0.623$). } \label{sketch_ring_coupler}
\end{figure}
As a simple model, we consider a pair of two concentric step-index
ring waveguides shown in Fig.~\ref{sketch_ring_coupler}(a). The two
concentric cores and the substrate are made of two different
materials with the refractive indices $n_1$ and $n_0$, and both
possessing a nonlinear Kerr response, so that the scalar optical
field $\psi(r,\phi,z)$ propagating in the $z$-direction may be
described by the normalized equation,
\begin{equation}
i\frac{\partial \psi}{\partial z}+\nabla_{\perp}^2 \psi + \left[V(r)
+|\psi|^2\right]\psi=0,
\end{equation}
where $\nabla_{\perp}^2$ is the Laplace operator, and $V(r)$ is the
external potential of the double-ring coupler. Since we are
interested in ring-shaped fields with nonzero angular momentum, the
indices and dimensions of the coupler were chosen so that, for the
linear regime, the coupler supports the usual symmetric and
antisymmetric vortex-type modes [plotted in
Fig.~\ref{sketch_ring_coupler}(b), solid]. Also, each of the
waveguides, when considered separately, supports the lowest-order
vortex-type radial mode [Fig.~\ref{sketch_ring_coupler}(b), dashed].
Those modes have the form,
\begin{equation}\label{input_field}
\psi(r,\phi,z)=u(r)\exp(i \ell \phi)\exp(i \beta z),
\end{equation}
where $u(r)$ is the radial profile of the corresponding mode,
$\beta$ is its propagation constant, and $\ell$ is the winding
number that we fix as $\ell=1$.

\begin{figure}[htbp]
  \centerline{
        \scalebox{0.3}
        {
        \includegraphics{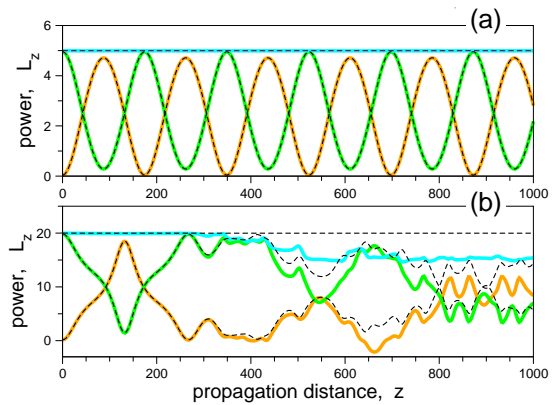}
        }
  }
\caption{Power (dashed lines) and angular momentum (continuous lines) in 
normalized units vs. propagation distance for outer-ring excitation. The 
individual values for each core are plotted, together with the total values. 
(a) simulation for P=5; (b) simulation for P=20. } 
\label{switching_curves_ext}
\end{figure}

\begin{figure}[htbp]
    \centerline{
        \scalebox{0.3}
        {
      \includegraphics{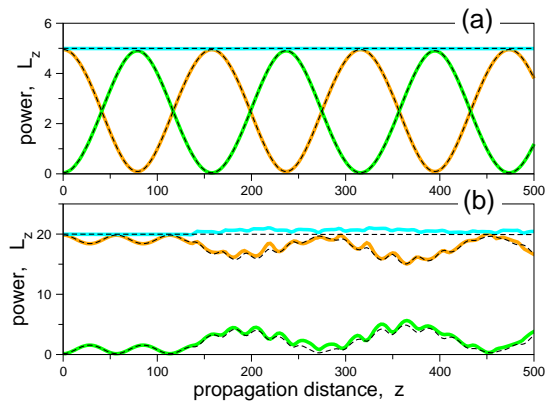}
        }
    }
\caption{Power (dashed lines) and angular momentum (continuous 
lines) vs. propagation distance for inner-ring excitation. The individual values 
for each core are plotted together with the total values. (a) simulation for 
P=5; (b) simulation for P=20.} 
\label{switching_curves_int}
\end{figure}

In our numerical simulations, we launch a vortex beam into one of
the ring waveguides. We chose the initial radial shape of the field
to be that of the single-waveguide mode
[Fig.~\ref{sketch_ring_coupler}(b), dashed] in order to assure a
good coupling to the ring-waveguide, although there would be no
problem to take some other shape like Gaussian. The field is also
initially scaled to establish the desired power and then propagates
in a nonlinear regime. At each value of $z$, we characterize the
vortex beam by calculating the beam power, $P=\int |\psi|^2 ~r dr
d\phi$, and its angular momentum, $L_z=\textrm{Im}\{\int
\psi^*\partial_\phi \psi ~r dr d\phi\}$, which are monitored for
each waveguide. To do so, we consider a circular boundary just half
way between both the cores and use all the points located in the
inner part of this boundary to calculate the power and momentum for
the inner core, and those in the external part to calculate those
values for the external core.
\begin{figure}[htbp]
    \centerline{
        \scalebox{0.35}
            {\includegraphics{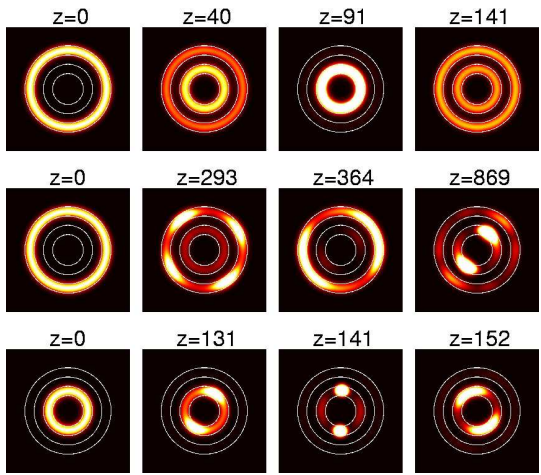}}
    }
\caption{Distribution of the light intensities in the coupler for
different propagation distance $z$, for three different regimes.
Top: external core excitation for the initial power $P=5$. Center:
external core excitation, at $P=20$. Bottom: internal core
excitation, at $P=20$.} \label{images_simulation}
\end{figure}
In Fig.~\ref{images_simulation}, we show the intensity images of the
coupler for three different switching scenarios. Also, for both
internal and external ring excitations, in
Fig.~\ref{switching_curves_ext} and Fig.~\ref{switching_curves_int}
we plot the power and angular momentum for each waveguide together
with their total values. Both magnitudes are in normalized units and
on the same scale. In fact, according to the functional form of the
input field (Eq.~(\ref{input_field})), they are initially
proportional,
\begin{equation}\label{relation_momentum_power}
L_z=\textrm{Im}\left\{ \int {\psi^* \partial_{\phi} \psi}=
\ell \int {u(r)^2} r dr\right\} = \ell P.
\end{equation}
The top row in Fig.~\ref{images_simulation} shows the case of the
external core excitation with a low power beam. A periodic coupling
between both cores takes place and the beam remains stable. If power
is increased (center row), the azimuthal instability breaks the beam
into a number of the fundamental solitons after some propagation
distance. The larger the input power the shorter the propagation
distance at which the breakup occurs. The resulting fundamental
solitons present an oscillating width and rotate inside the ring due
to the angular momentum, with an angular velocity that is larger as the
the solitons width is smaller. If the input power is further increased, the
fundamental solitons collapse as usual happens in the Kerr media. It is
well known that for the nonlinear media with the focusing nonlinearity,
solitons are unstable so that they either collapse or spread out. Due to the
presence of the waveguides the spreading is stopped and so for low powers
the beam remains stable.

Due to the proportionality relation between the momentum and power
(see Eq.~(\ref{relation_momentum_power})), and the fact that we chose
$\ell=1$, the power and angular momentum have the same value
meanwhile the beams remain unbroken (notice coincident thin and thick lines in
Fig.~\ref{switching_curves_ext} and
Fig.~\ref{switching_curves_int}). Nevertheless, it is notable that
this parallel behavior diverges from the point where the beams
breakup occurs due to the instabilities (notice a separation of the
thin and thick lines), revealing differences in the momentum
coupling. Besides, the total angular momentum is not longer conserved as is
shown in the plots.
\begin{figure}[htbp]
    \centerline{
        \scalebox{0.3}
        {
      \includegraphics{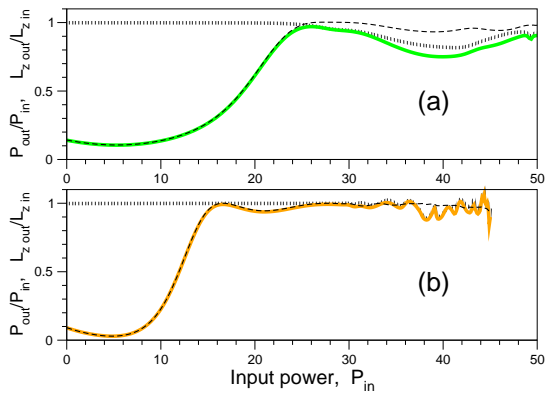}
        }
    }
\caption{Switching curves for the power (dark dashed lines) and
angular momentum (light continuous lines) for (a) outer-waveguide
excitation, and  (b) inner-waveguide excitation. Also shown is the
total angular momentum (dotted line).} \label{switching_curves}
\end{figure}
Another interesting fact is the difference in suppression of
coupling (switching) between internal and external ring excitation.
When the external-ring core is excited, since it has a larger area, a higher
power density is coupled into the inner core, and the beam breaks for a lower
input power than the necessary to complete the switching. This is
evident from Fig.~\ref{images_simulation}(center row) where it is shown that
after the beam breakage, coupling is still possible for a long enough
propagation distance. To account for the switching property, we show the
switching curves in Fig.~\ref{switching_curves}, for power and angular
momentum. They were built launching a single-ring field into one of the cores,
simulating the propagation for half a beating length and measuring the output
power at the same core (the plot is the ratio between output and input power).
The beating length is defined as the necessary distance to
perform a complete coupling cycle (power transfer to the second core and back
to the first one), and was obtained by simulation in the linear regime. For
the parameter values considered the half beating length resulted in z=73.75
and z=73.85 for internal and external ring excitation respectively.

Note that due to the asymmetry of the coupler (rings of different
diameter), a complete transfer of power does not take place for the
linear regime. That is why raising the power produces an initial
increase of the coupled power and the switching curve presents an
decreasing behavior at low powers. Also, for external core
excitation, Fig.~\ref{switching_curves}(a), the triggering is more
steady and takes place at higher powers than for internal core
excitation (b). This is due to the lower density of power in the
external core because of its larger area.

The separation between the power and angular momentum curves denotes
the point where the azimuthal instability breaks the beam before it
propagates half a beating length. It happens clearly after switching
took place in the case of internal excitation, but slightly before
that point for external excitation. The curves reach an end-point
when the power is so high that collapse takes place before the beam
propagates for half a beating length.

In conclusion, we have suggested a novel design of a nonlinear
optical coupler for switching the angular momentum of light. We have
studied numerically the operation of such a coupler and demonstrated
that a low-power optical vortex launched into one of the
ring waveguides transfers periodically between both rings with a perfect
periodic change of both beam power and its angular momentum. From the study
of the nonlinear regime (larger power) of the coupler we draw the conclusion
that the internal excitation is more effective since the slope of the switching
curve is larger and switching is triggered well before the beam breaks up due
to the azimuthal instability. For outer-ring excitation, however, it is still
possible to perform the switching for powers low enough to avoid the beam
collapse.

JRS acknowledges the Ram\'on y Cajal contract from the Ministerio de
Educaci\'on y Ciencia of Spain and a support from Xunta de Galicia
(project PGIDIG06PXIB239155PR). YK acknowledges a support from the
Australian Research Council.

\end{document}